\documentstyle[12pt,epsf]{article}
\newcommand{\dd}{\mbox{d}}
\newcommand{\etal}{et~al.}
\def\lesssim{\mathrel{\hbox{\rlap{\hbox{\lower4pt\hbox{$\sim$}}}\hbox{$<$}}}}

\begin{document}
%
%
\begin{center}{\bf Constraints on the Gamma-ray Burst Luminosity Function from
PVO and BATSE Observations} \\
Andrew Ulmer\footnote{andrew@astro.princeton.edu} \\
Princeton University Observatory,
       Princeton, NJ~08544-1001  \\
Ralph A.M.J. Wijers\footnote{ramjw@mail.ast.cam.ac.uk} \\
Institute of Astronomy, Madingley Road,  Cambridge CB3 0HA, UK \\
Edward E. Fenimore\footnote{efenimore@lanl.gov} \\
Los Alamos National Laboratory, MS D436, Los Alamos, NM 87545
\end{center}
%
%
%
%
   \begin{abstract}

We examine the width of the gamma-ray burst luminosity function through
the distribution of GRB peak fluxes as detected by the Pioneer Venus
Orbiter (PVO) and the Burst and Transient Source Experiment (BATSE).  The
strength of the analysis is greatly enhanced by using a merged catalog of
peak fluxes from both instruments with good cross-calibration of their
sensitivities. The range of peak fluxes is increased
by approximately a factor of 20 relative to the BATSE catalog. Thus, more
sensitive investigations of the $\log N-\log P$ distribution are possible.
We place constraints on the width of the luminosity function of gamma-ray
bursts brighter than the BATSE completeness limit by comparing the
intensity distribution in the merged catalog with those produced by a
variety of spatial density and luminosity functions.  For the models
examined, $90\%$ of the {\em detectable\/} bursts have peak luminosities
within a range of 10, indicating that the peak luminosities of gamma-ray
bursts span a markedly less wide range of values than many other of their
measurable properties.
We also discuss for which slopes of a power-law
luminosity function the observed width is at the upper end of the constrained
range.
This is important in determining the power-law slopes
for which luminosity-duration correlations could be important.

\end{abstract}
\begin{center}
\em{subject headings:}
gamma-rays: bursts --- methods: data analysis, statistical
\end{center}

%
%
\section{Introduction}

The importance of understanding the gamma-ray burst luminosity function
has recently been underscored by works (Brainerd 1994, Mao \& Yi 1994)
discussing luminosity-timescale correlations within a Galactic corona
model as possible producers of the detected time-stretching between weak
and strong bursts (Norris \etal\ 1994).
In the cosmological scenario, if cosmological
redshifts {\em alone} are to account for the time-stretching,
then either the luminosity function must be relatively narrow or
luminosity-timescale correlations must be small.
The luminosity function, therefore, likely plays an important role in
either a Galactic or a cosmological scenario.

Determining the luminosity function of GRBs is difficult because there is
no direct information about the distance to any individual burster. As a
result, the luminosity function can only be examined as an inverse problem
involving the detected intensities and the inferred spatial distribution.
Likewise, an understanding of the spatial distribution is hampered by
uncertainties in the luminosity function.  Assuming the luminosity
function is independent of distance, we have
\begin{equation}
\label{npmax}
N(P^\prime > P)= \int_{L_{1}}^{L_{2}} {\rm d}L \int_{0}^{D(L)}
{\rm d}R 4\pi R^2\rho(R)f(L)
\end{equation}
where $N(P^\prime>P)$ is the number of bursts detected with peak flux
greater than $P$, $f(L)$ is the luminosity function, $\rho(R)$ is the
spatial density function (with $R$ the distance from us), and $D(L) =
\sqrt{L/4\pi P}$.  In cosmological scenarios, the expression is more
complicated due to redshifts and possible evolutionary effects.  While it
is not possible to uniquely deconvolve the luminosity and spatial
distributions with only the $\log N-\log P$ information, we can study
limits on the luminosity function given certain assumptions about the
general form of the spatial density function.

Previous studies of the luminosity function have analyzed $\log N-\log P$
distributions from the BATSE catalog of Fishman \etal ~(1994).  They have
focused on power-law luminosity functions as a first approximation since
the parameter space is relatively small for this case and a power law is
often found to be a good first approximation for the luminosity function
of astronomical objects.  Emphasis in these studies has been the width of
the luminosity function:  are the peak luminosities of gamma-ray bursts
approximately standard candles or do they span a wide range? One should
bear in mind that there is no information on bursts that are below the
instrument detection threshold. Since the luminosity functions assumed or
inferred from fits generally rise towards low luminosities, there could be
a large population of intrinsically weak gamma-ray bursts that are never
detected and are therefore missed in our analysis and in other analyses of
the same problem.  When we speak of the luminosity function here, we mean
the luminosity function of {\em detected\/} bursts (which we call the
observed luminosity function) unless otherwise noted.  Ulmer and Wijers
(1995, hereafter UW95) concluded that the BATSE distribution did not
contain enough bright bursts in the sample to strongly limit the location
of the bend in the $\log N-\log P$ curves.  Consequently, they found that
a typical limit to the width of the observed luminosity function was
substantial but limited (20--40), but there was a heavy tail of very large
widths ($>$100) in the distribution that could not be ruled out with
confidence.  Horack, Emslie, and Meegan (1994) found stronger
constraints on the width using a novel technique which
has the advantage of being insensitive to the spatial density distribution
(Horack \& Emslie 1994).
Their values appear consistent with those of UW95 and the work presented here,
and any discrepancies are likely due to different definitions of the width
and a different sample of bursts.
At the very least, these works are indications
that the observed luminosity function may be narrow.

\section{Improved Analysis}

The method followed here is similar to that of Ulmer and Wijers (1995,
UW95).  In order to improve the power of the analysis of the width of the
gamma-ray burst distribution, we extend the available data set to include
more bursts on the bright side of the bend in $\log N - \log P$
by combining the data from the BATSE catalog and the PVO catalog.
Because the position of the bend and correction for incompleteness are
crucial to a proper treatment of the issue, we need to combine them with
extreme care.  A merging of the BATSE and PVO catalogs was previously
performed in order to study the shape of the $\log N - \log P$
distribution and estimate the intrinsic luminosity of gamma-ray bursts in
standard cosmological scenarios (Fenimore \etal ~1993). The merging
procedure involves determining a common energy band (100-500 keV), trigger
timescale (250\,ms or 1000\,ms), and detection efficiencies; it is
described in Fenimore \etal ~(1993). We use a combination of bursts from
250 and 1000\,ms trigger times due to constraints in the PVO database.
Relevant details of the detection
efficiencies over the range of burst intensities are also discussed by in
't Zand \& Fenimore (1993).  Because of the long exposure and high
fraction livetime of PVO, it would take BATSE approximately 30 years to
achieve the dynamic range of the combined PVO/BATSE distribution, so
analyzing combined data sets will always remain more powerful than
analyzing data from just one instrument.  Here we utilize peak fluxes in
photons~cm$^{-2}$~s$^{-1}$ for 99 BATSE events brighter than
1.0\,photons~cm$^{-2}$~s$^{-1}$ and 139 PVO events brighter than
20\,photons~cm$^{-2}$~s$^{-1}$, with the peak fluxes defined for 250\,ms
time bins.

With the data sets on a common flux scale, we can extend the analysis from
UW95.  Briefly, the method entails comparing the $\log N - \log P$
distribution with a theoretical distribution determined by convolving
luminosity functions and spatial density functions as in equation 1.  For
this analysis, we use spatial density distributions of the form
\begin{equation}
\label{halo}
\rho(R) = \frac{\rho_{\rm o}}{1+(R/R_{\rm c})^{\alpha}},
\end{equation}
where $R$ is distance from the Earth,
and $R_{\rm c}$ is the core radius (e.g. $10^4$ AU or 20 kpc),
and $\alpha$
parameterizes the density fall off rate. This choice reflects the fact
that we know the density of gamma-ray bursts to be constant nearby, and
falling beyond some characteristic distance (either because the density
truly decreases or, in cosmological models, due to decrease of the volume
element at high redshift). For different values of $\alpha$ these
distributions cover a wide range of spatial density functions, from those
which fall only slowly to zero to those which abruptly stop at $R_{\rm
c}$.  The luminosity function is parametrized as a truncated power law,
\begin{equation}
\label{pw}
f(L) = C L^{-\beta}  \hspace{.5 in} L_{1} < L < L_{2}.
\end{equation}

We determine the best fit function parameters via maximum likelihood
fitting of the data. This technique has the advantage that no information
is lost as a result of binning.  The likelihood function is
\begin{equation}
f = \left(\left.\prod_{i=1}^{N_{\rm PVO}} -
\frac{\dd N_{\rm PVO}(P^\prime > P; X)}{\dd P}\right]_{P=P_i}\right)
\left(\left.\prod_{j=1}^{N_{\rm BATSE}} -
\frac{\dd N_{\rm BATSE}(P^\prime > P; X)}{\dd P}\right]_{P=P_j}\right),
\end{equation}
where $N(P^\prime > P; X)$, given in eq. 1, is the number of bursts with
peak flux greater than $P$, and $X$ are parameters of the spatial density
and luminosity functions. (Note that one of ($R_{\rm c}$, $L_1$, $L_2$) is
redundant, since the fit function depends on these three through the
two fluxes $L_{1,2}/4\pi R_{\rm c}^2$.) In this expression, the
theoretical distributions differ between the two instruments for
the same $X$, because the distribution must be truncated at the
sensitivity limit of the instrument in question, and normalized to unit
integral above that limit.

For a maximum likelihood set of parameters, we can calculate the width of
the {\it observed} luminosity function above the BATSE completeness
limit.  We take the 90\% width to be $L_{95\%}/L_{5\%}$. The luminosities
$L_{95\%}$ and $L_{5\%}$ are such that the intervals $(L_1,L_{5\%})$ and
$(L_{95\%},L_2)$ each contain $5\%$ of the total number of {\it detected}
bursts:
\begin{equation}
\label{width}
0.05 = \frac{\int^{L_{5\%}}_{L_{\rm min}} dL \int_{0}^{D_{\rm max}(L)}dR
4\pi R^2\rho(R)f(L)}{\int_{L_{\rm min}}^{L_{\rm max}} dL
\int_{0}^{D_{\rm max}(L)}
dR 4\pi R^2\rho(R)f(L)},
\end{equation}
\begin{equation}
\label{width1}
0.05 = \frac{\int_{L_{95\%}}^{L_{\rm max}} dL \int_{0}^{D_{\rm max}(L)}dR
4\pi R^2\rho(R)f(L)}{\int_{L_{\rm min}}^{L_{\rm max}} dL
\int_{0}^{D_{\rm max}(L)} dR 4\pi R^2\rho(R)f(L)}
\end{equation}
where $D_{\rm max}(L) = \sqrt{L/4\pi P_{\rm min}}$ and $P_{\rm min}$ is
1.0 photon~cm$^{-2}$~s$^{-1}$.  Confidence intervals on the maximum
likelihood parameters are determined by bootstrap resampling of the data.

\section{Constraints on the Width}

Limits on the width of the observed luminosity function are shown in
figure 1.  With the large dynamic range of the combined BATSE and PVO
catalog, the bend and $-3/2$ part of the $\log N - \log P$ distribution
are well sampled. It is possible to place a limit of about 10 on the range
in which $90\%$ of the bursts with peak flux greater than 1.0
photon~cm$^{-2}$~s$^{-1}$ are contained.  While this result depends
somewhat on the assumed forms of the spatial density function and the
luminosity
function, the result that the observed width is relatively small seems
firmly established, for all those cases where the actual form of the
spatial density function is not widely different from what we assumed.
This finding
offers a significant constraint on our understanding of gamma-ray bursts,
because it appears that within a range of less than ten in intrinsic
luminosity, gamma-ray bursts are able to produce their wide range of
durations (5 orders of magnitude) and fluences, as well as a collection of
time histories that span a range of qualitatively different forms.  We
note again that it is the width of the observed luminosity function that
is limited and not necessarily the width of the intrinsic luminosity
function: there may be many gamma-ray bursts which have low enough
luminosities that their intensities are below the current detector
sensitivities.

We also illustrate the relationship between the width of the luminosity
function and the power-law indices of the luminosity function and spatial
density function in figure 2.  For large $\beta$ the bursts are
essentially standard candles because the weak end of the luminosity
function dominates; for small $\beta$ this is again the case, with the
bright end dominating.  As discussed in Ulmer and Wijers (1995), the
luminosity functions can only be effectively wide if $1\le\beta\le2.5$; indeed,
the fitted widths are seen to be large only for values of $\beta$ in that
range.  One can also use this figure for qualitative analysis of power law
luminosity functions. For instance, for models which involve some type of
asymptotic $1/r^2$ dependence of the spatial density distribution (dark
matter halo), the figure shows that for power law indices between about 2
and 3, luminosity functions may have observed widths around 10. For
luminosity functions which are much steeper or much shallower, the bursts
would behave as standard candles even if their intrinsic range of
luminosities (i.e., $L_2/L_1$ in eq.~3) were quite large.

It is virtually impossible to place interesting restrictions on
the range of acceptable values of
$\alpha$ from the data. This is due to an unavoidable ambiguity in fitting
a $\log N - \log P$ curve with a $-3/2$ slope at the bright end and a
shallower one at low fluxes (e.g. UW95). One can regard the
objects as standard
candles (i.e.\ $\beta<1$ or $\beta>2.5$ or $L_1\simeq L_2$) and consider the
faint-end ${\rm d}\log N/{\rm d}\log P$ to directly reflect the density as
a function of distance for large distances, which implies ${\rm d}\log
N/{\rm d}\log P= (\alpha-3)/2$ in our case.
It is possible to place lower bounds on $\alpha$ no higher
than 1.4, the value for which the slope in $\log N - \log P$ is completely
provided by the spatial spatial density function.
Or one can choose a strong
density cutoff, $\alpha>3$, in which case the faint-end slope directly
reflects the slope of the luminosity function, ${\rm d}\log N/{\rm d}\log
P = 1-\beta$. This behavior is seen in figure 2. All fits in the displayed
range of $\alpha$ are acceptable, but the preferred value of $\beta$
shifts from 1.8, the value that directly fits the faint-end ${\rm d}\log
N/{\rm d}\log P$, at large $\alpha$ to a standard-candle like value of
about 2.5 at small $\alpha$.  This is also visible in figure 1 as a
decrease in the width of the luminosity function towards low values of
$\alpha$.

   \section{Discussion and Conclusion}

We have derived the strongest constraints yet on the range of intrinsic
peak luminosities of gamma-ray bursts that have observed peak fluxes above
the BATSE completeness limit: 90\% of these must fall within a range of
10.
It is surprising that within a range of less than ten in intrinsic
luminosity, gamma-ray bursts produce a wide range of fluences,
durations spanning 5 orders of magnitude, as well as time histories with
strikingly different forms.
The constraint was derived under the simplifying assumption that
the luminosity function does not depend on distance, and that the spatial
density and luminosity functions can parametrized in a certain way. For
Galactic coronal or Oort cloud models, these constraints should be quite good.
For
functional forms that greatly differ from what we assumed, for example a
two-population luminosity function, the limits will be somewhat
different.  If gamma-ray bursts are at cosmological distances, our
results are not directly applicable because we did not account for
redshift and space curvature in our computation of the expected flux
distribution.  Also, the assumption that the luminosity function is
independent of distance is less likely to be valid in the cosmological
case, because most known objects show marked evolution between redshift 0
and 1.

The luminosity function of detected bursts can only have widths at the upper
end of the constrained range of widths if $\beta$, the
power-law slope of the luminosity function (see eq. 3), is in the range
1--2.5.  If a luminosity-duration correlation is used to
explain the time-stretching phenomenon in dim bursts found by Norris
\etal\ (1994), the effects of such correlations can only be important
if a range of luminosities are observed. For such models,
$\beta$ is likely to be between 1 and 2.5, so that the
observed width will be as large as possible. However, it is conceivable that
with strong correlations and appropriate spatial density and intrinsic
luminosity functions, the observed luminosity function need not
have a width of more than a couple to produce strong time-stretching effects.
Relativistic
beaming has been proposed as a way of introducing such a correlation
(Brainerd 1994).  For the case of an infinitely narrow beam, the
luminosity function produced by different viewing angles has a slope of
approximately 1.25 (Yi, 1993).  However, the relativistic beam sizes are
finite and in all likelihood larger than $1/\gamma$ (Mao \& Yi, 1994), and
the luminosity functions are nearly standard candles for fixed $\gamma$.
While the viewing angle for these models is probably unimportant, the
$\gamma$ of the beam may vary between bursts and produce time-stretching
effects.  We suggest that for models in which a variation of $\gamma$
among bursts is used to introduce the required luminosity-duration
correlation, the resulting luminosity function should again have
$1\lesssim\beta\lesssim 2.5$.

Acknowledgements:

We thank B. Paczy{\'n}ski and J. Horack, the referee, for helpful comments.
AU is supported by a NSF Graduate Research Fellowship. RAMJW is supported
in part by a Compton Fellowship, grant GRO/PFP-91-26, and in part by a
PPARC fellowship. Part of this work was
also supported by grant NAG5-1901 and the CGRO Guest investigator program.

\newpage



\begin{center}
References:
\end{center}





\par\noindent\hangindent=3pc\hangafter=1
Brainerd, J.J., 1994, ApJ., 428, L1

\par\noindent\hangindent=3pc\hangafter=1
Fenimore, E.~E., Epstein, R.~I., Ho, C., Klebesadel, R.~W., Lacey, C., Laros,
  J.~G., Meier, M., Strohmayer, T., Pendleton, G., Fishman, G., Kouveliotou,
  C., \& Meegan, C., 1993, Nat, 366, 40.


\par\noindent\hangindent=3pc\hangafter=1
Fishman, G.J., \etal , 1994, ApJS, 92, 229

\par\noindent\hangindent=3pc\hangafter=1
Horack, J.~M., \& Emslie, A.~G., 1994, ApJ, 428, 620

\par\noindent\hangindent=3pc\hangafter=1
Horack, J.~M., Emslie, A.~G., and Meegan, C.~A., 1994, ApJ, 426, L5

\par\noindent\hangindent=3pc\hangafter=1
in 't Zand, J.J.M. \& Fenimore, E. E., 1993,  in
{\it 2nd Huntsville Gamma-ray Burst Workshop} (eds
Fishman, G., Brainerd, J., \& Hurley, K.), 692


\par\noindent\hangindent=3pc\hangafter=1
Mao, S. \& Yi, I. 1994, ApJ, 424, L131


\par\noindent\hangindent=3pc\hangafter=1
Norris, J. P., Nemiroff, R.J. Scargle, J.D., Kouveliotou, C., Fishman, G.J.,
Meegan, C.A., Paciesas, W.S., \& Bonnel, J.T. 1994, ApJ, 424, 540

\par\noindent\hangindent=3pc\hangafter=1
Ulmer, A. \& Wijers, R.A.M.J. (UW95) 1995, ApJ., in press

\par\noindent\hangindent=3pc\hangafter=1
Yi, I. 1993, Phys. Rev. D, 48, 4518

\newpage

{\bf
  \begin{center}
     Figure captions\\[3em]
  \end{center}
}

\begin{figure}[h]
 \caption{Constraints on the width of the observed luminosity function are
shown for a large range a spatial density functions (see Eq. 2).
The solid line corresponds to
$L_{95\%}/L_{5\%}$ at 90$\%$ confidence, where $90\%$ of the bursts
observed by BATSE (with peak flux greater than 1.0
photon~cm$^{-2}$~s$^{-1}$) had  luminosities between $L_{95\%}$ and $L_{5\%}$.
The dashed line shows $L_{90\%}/L_{10\%}$ at $90\%$ confidence.}
\end{figure}

\begin{figure}[h]
 \caption{This figure illustrates qualitative dependences of the width of
the luminosity function on the index of the luminosity function (eq. 3) and
the spatial density fall off rate.
Maximum likelihood fits to bootstrap resamplings
of the data are depicted for a series of fixed values of the spatial density
function (with 1000 resamplings for each exponent value).
The size of the crosses indicates the width of the luminosity
function ($L_{95\%}/L_{5\%}$),
with the smallest corresponding to a width of 3 or less and
the largest corresponding to a width of 15 or more (which occurs in only
a small fraction of the simulations.)}
\end{figure}

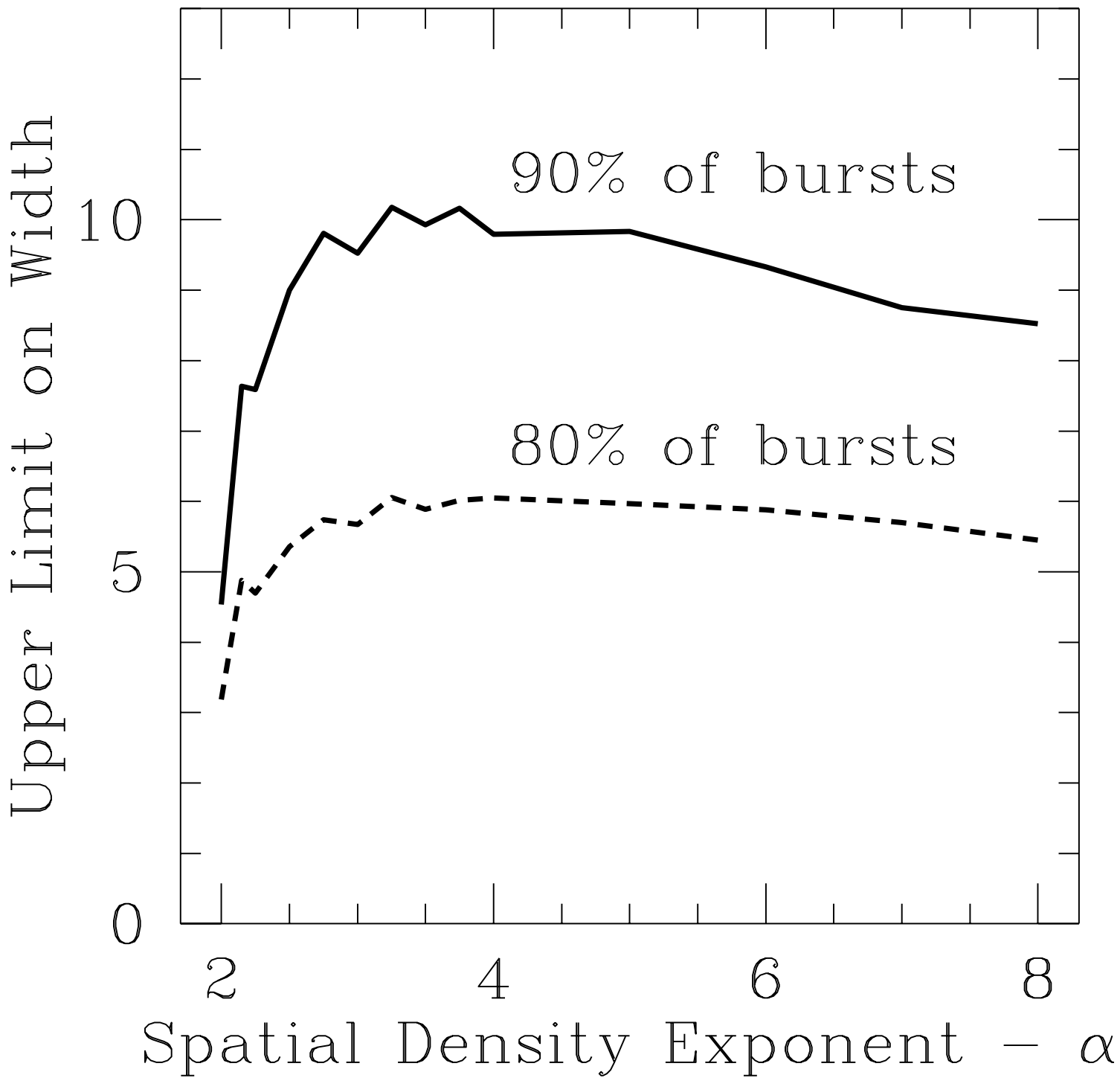
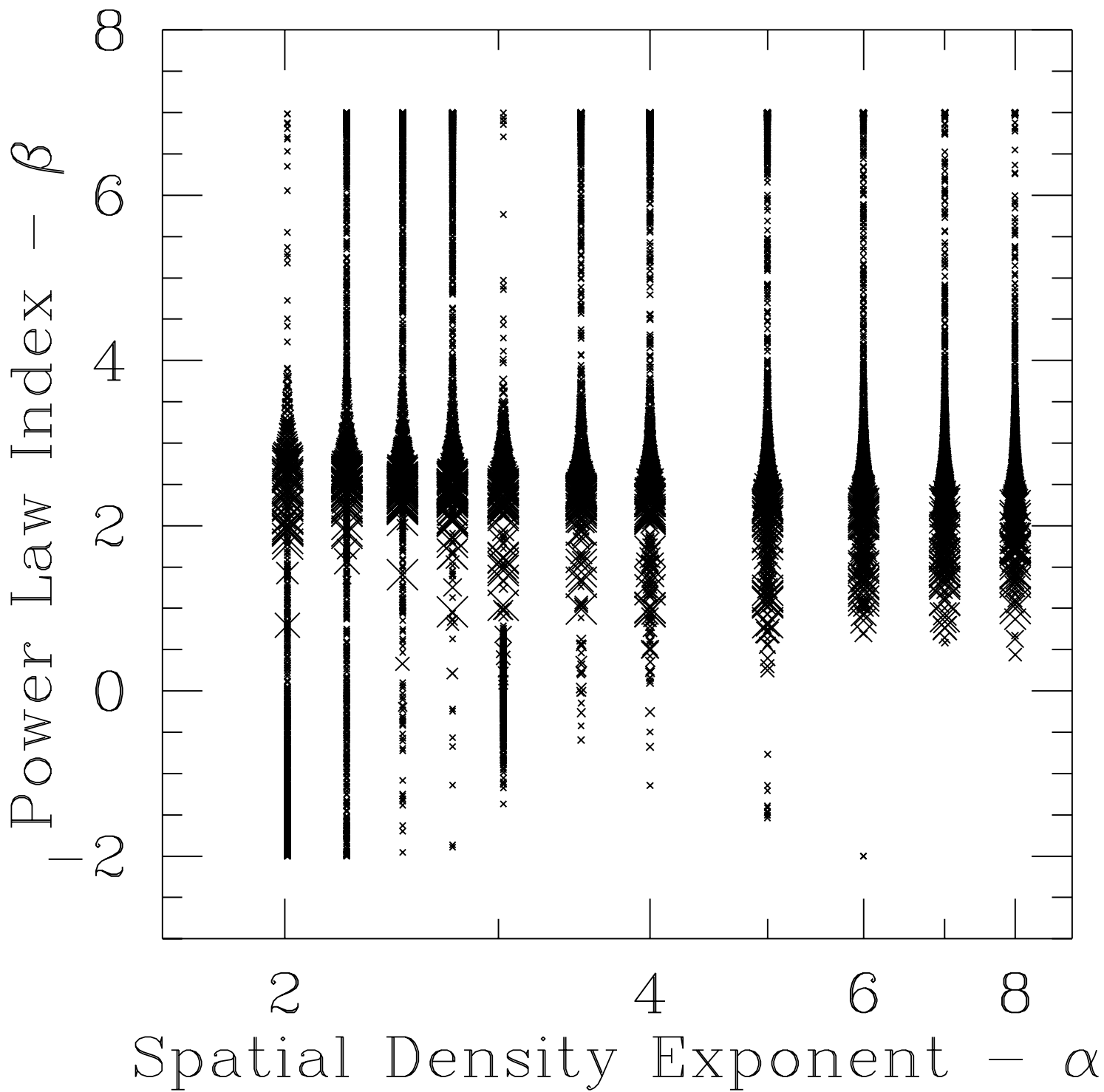
\end{document}